\documentclass[aps,twocolumn,a4paper,10pt]{revtex4-1}
\usepackage{mathrsfs}
\usepackage{graphicx}
\usepackage{latexsym}
\usepackage{amsmath}
\usepackage{amssymb}
\usepackage{textcomp}
\usepackage{amsbsy}
\usepackage{graphics}
\usepackage{epstopdf}

\begin{document}
 \title{\large \bf Higher rank antisymmetric tensor fields in Klebanov-Strassler geometry}
\author{Ashmita Das}
\email{tpad@iacs.res.in}
\author{Soumitra SenGupta}
\email{tpssg@iacs.res.in}
\affiliation{Department of Theoretical Physics,\\
Indian Association for the Cultivation of Science,\\
2A $\&$ 2B Raja S.C. Mullick Road,\\
Kolkata - 700 032, India.\\}

\begin{abstract}
 In string theory, higher rank antisymmetric tensor fields appear
as massless excitations of closed strings. 
Till date there is no experimental support in favour of their existence.
In a stringy framework, starting 
from a warped throat-like Klebanov-Strassler geometry,
we show that all the massless higher rank antisymmetric 
tensor fields are heavily suppressed due to the background fluxes
leading to their invisibility in our Universe.
\end{abstract}
\maketitle
\section*{Introduction}
A complete theoretical framework for quantisation of gravity and unification 
of four fundamental interactions is 
a long standing problem for decades.
Till date many theories have been proposed in this regard. Among 
different proposals, string theory is 
elegantly designed to search for a complete theoretical understanding 
of these basic aspects. 
It is well known that in the low energy effective theory,
superstring theories contains many massless higher rank antisymmetric
tensor fields as different excitations of string.
 Among such antisymmetric tensor fields,
 the rank two antisymmetric KR field has some special features in the context of the 
spacetime geometry. It has been argued that the corresponding rank three antisymmetric
field strength tensor $H_{\mu\nu\lambda}(x)$,
can be interpreted as the antisymmetric connection (known as torsion)
of the underlying spacetime \cite{ssgpm}. \\
A long awaited answer for the question that why torsion 
has no visible effects on our Universe (known as visible 3-brane)
while curvature has dominating impacts, has been addressed by the author of \cite{ssgkr}
in the background of the Randall-Sundrum (RS) \cite{RS}
warped spacetime geometry.  Subsequently it was shown by the same authors that 
all the other higher rank antisymmetric 
tensor gauge fields in general are also suppressed on our brane due to the warped
character of RS geometry \cite{ssg_higherrank}.\\
In this context it is worthwhile to re-examine this feature in
the backdrop of a warped geometry originating from a stringy compactification.\\
Here we consider a special kind of  warped compactification of
10-dimensional type IIB supergravity in presence of background fluxes,
which is popularly known as Klebanov-Strassler (KS) geometry. 
Analysing the behaviour of various higher rank 
antisymmetric tensor gauge fields in the background of
Klebanov-Strassler geometry we try to address the 
issue of non-visibility of these gauge fields in our observable Universe.

\section{\small Klebanov-Strassler geometry: A brief description}\label{KS geometry}
In string theory Klebanov-Strassler warped throat-like 
spacetime geometry \cite{KS_2000} plays a crucial role in addressing many 
fundamental questions. Many authors in recent past have shown that the presence of warped throat 
region with D3 and anti-D3 branes can offer an explanation 
to different inflation and slow-roll models
 \cite{verlinde,kachru,buchel,baumann1,bean,shiu,kofman,baumann2,baumann3,kachru_1,kallosh}.
Moreover this warped throat-like deformation of spacetime produces a geometrical resolution
to the gauge hierarchy problem where all the low energy phenomena are confined within a highly red shifted region.
This region is identified as infrared brane (IR brane), located near the tip of the throat \cite{kachru_hierarchy}.\\
KS geometry is an example of a warped compactification of 10-dimensional type IIB supergravity in presence of 
background fluxes \cite{KS_2000}. The resulting spacetime geometry is the warped
product of the 4-dimensional Minkowski spacetime with 6-dimensional Calabi-Yau (CY) orientifold.
The origin of KS geometry is described in \cite{KS_2000,verlinde,kachru,baumann3,kachru_hierarchy,kachru_1}
 by considering  a compactification of string theory on $AdS_{5}\times X_{5}$ where
$X_{5}$ is the Einstein manifold in five dimensions.
In case of KS geometry the above compactification reduces the background spacetime to,
$AdS_5 \times T^{1,1}$ where $T^{1,1} \sim S^2 \times S^3$ : a 5-D Einstein spacetime.
Considering the radial direction present in $AdS_5$, the resulting
 six dimensional internal space has an end in the form of a tip
 which is joined to an $S^3$ of finite size and 
a warping can be produced by the background fluxes, making the 
spacetime geometry as a deformed conifold. The background fluxes are \cite{kachru}:
\begin{equation}
\frac{1}{2(\pi)^2\alpha^{'}}\int_{A} F=M~, ~~~~~~~\frac{1}{2(\pi)^2\alpha^{'}}\int_{B} H=-K\label{Acycle}
\end{equation}
Here A cycle is in the form of $S^3$ at the tip of the conifold and B cycle is
the Poincare dual three-cycle. $M, K >>1$ are integers. 
In this setup the spacetime coordinates are, 4-D Minkowski coordinates and
radial coordinate $r$ symbolises the distance 
from the tip of the conifold. In addition five more angles define the orientifold $T^{1,1}$.
The metric ansatz for this KS geometry can be written as \cite{kachru},
\begin{equation}
ds^2=h^{-1/2}\eta_{\mu\nu}dx^{\mu}dx^{\nu}+h^{1/2}(dr^2+r^2 ds^{2}_{T^{1,1}})\label{ksmetric}
\end{equation}
where, 
\begin{equation}
h(r)=\frac{27\pi}{4r^4}\alpha^{'2}g_s M\big(K+g_sM \bigg(\frac{3}{8\pi}+\frac{3}{2\pi}
\rm{ln}\bigg(\frac{r}{r_{max}}\bigg)\bigg)\bigg)\label{warpfactor2}
\end{equation}
Defining, 
\begin{equation}
R^4=\frac{27}{4}\pi g_sN\alpha^{'2},~~~~~~~~~~~~~~N\equiv MK\label{stringparameters1}
\end{equation}
and neglecting the second and the logarithmic terms on the
right side of eqn(\ref{warpfactor2}) \cite{kachru},
the warp factor reduces to 
\begin{equation}
 h(r)=\frac{R^4}{r^4}\label{warpfactorKS}
\end{equation}
We restrict our analysis in the
region  $r_0<r<r_{max}$ of the  $\rm{AdS}_5$.
 It is shown in 
\cite{KS_2000,verlinde,kachru,baumann3,kachru_hierarchy,kachru_1},
 that the slicing of the $AdS_5$ region within $r_0$ and $r_{max}$
is a consequence of the 
compactification of the spacetime where away from the
region $r_0<r<r_{max}$, one obtains a geometry which is 
significantly dissimilar from the $AdS_5\times T^{1,1}$.
Therefore the presence of the warped throat region 
can distinctly be identified from the embedding spacetime geometry.
 In the ultraviolet (UV) limit {\it{i.e}}
$r\geqslant r_{max}$, the throat geometry becomes conical and smoothly connects with the CY orientifold.
At this connecting region the geometry deviates from $AdS_5 \times T^{1,1}$ 
so that $r_{max}$ can be projected as the ultraviolet cut-off in the Anti de-Sitter spacetime
\cite{KS_2000,verlinde,kachru,baumann3,kachru_hierarchy,kachru_1}. Simultaneously the throat terminates 
smoothly in the infrared (IR) region {\it{i.e}} the region of small $r$  
and in particular connects to an $S^3$ of finite size at $r=r_0$. 
The above description of KS geometry implies that
if we exclude $T^{1,1}$ from the background spacetime geometry, the remaining
$AdS_5$ is similar to RS warped geometry defined in five dimensions \cite{kachru}. Therefore 
like five dimensional RS scenario, 
we identify the two end points of this
trimmed $AdS_5$ spacetime as the location of UV brane/Planck brane (at $r=r_{max}$) 
and IR brane/standard model (SM) brane ({\it{i.e}} at $r=r_0$). We focus into the 
region $AdS_5$, which contains the 4-D Minkowski and a radial coordinate, 
and choose to ignore the angular coordinates defined on $T^{1,1}$ orientifold
from our discussion.
In the background of the KS throat geometry, the authors of \cite{kachru_hierarchy}, has 
evaluated the redshift at the tip of the conifold, located at $r_0$. 
The derived minimal redshift is dependent on the background fluxes $M$, $K$ and
can be written as,
\begin{equation}
 \frac{r_0}{R}=e^{-2\pi K/3g_s M}\label{warpfactor3}
\end{equation}
\section{\small Analysis of massless second rank KR tensor field in the 
background of the KS throat geometry}\label{KR field}
The $AdS_5$ metric in KS throat geometry can be written as,
\begin{equation}
 ds^2=h^{-1/2}(r)\eta_{\mu\nu}dx^{\mu}dx^{\nu}+h^{1/2}(r)dr^2\label{5dimmetric}
\end{equation}
The five dimensional action for the Kalb-Ramond field is \cite{ssgkr},
\begin{equation}
 S_5=\int d^5xH_{MNL}H^{MNL}\label{actionKR1}
\end{equation}
where each Latin and Greek index runs from $0$ to $4$
and $0$ to $3$ respectively.\\
As discussed in \cite{ssgkr}, we consider the source of torsion to be the rank-2
anti-symmetric Kalb-Ramond field $B_{MN}$. Torsion can be identified
with the rank-3 antisymmetric field strength tensor $H_{MNL}$ which is
related to the KR field as,
\begin{equation}
 H_{MNL}=\partial_{[M}B_{NL]}\label{fldtensor}
\end{equation}
Following the gauge symmetry, $B_{MN}\rightarrow B_{MN}+\partial_{[M}\Lambda_{N]}$,
we use the gauge condition $B_{4\mu}=0$.
The Kaluza-Klein (KK) decomposition for the remaining components of
KR field is now given as,
\begin{equation}
 B_{\mu\nu}(x,r)=\sum_{n=0}^{\infty}B^{n}_{\mu\nu}(x)
\frac{\chi^{n}(r)}{\sqrt{r_c}}\label{kkkr1}
\end{equation}
where $r_c$ is the distance between the UV and IR brane and $\chi^{n}(r)$
is the wavefunction for the KR field along the radial direction.
Substituting this in the 5-dimensional action and  integrating over the radial direction,
the four dimensional effective action turns out to be:
\begin{eqnarray}
 S^{(4)}_{H}=\sum_{n=0}^{\infty}\int d^{4}x
[\eta^{\mu\alpha}\eta^{\nu\beta}\eta^{\lambda\gamma}H^{n}_{\mu\nu\lambda}
H^{n}_{\alpha\beta\gamma}\nonumber\\
+3m_{n}^{2}\eta^{\mu\alpha}\eta^{\nu\beta}B^{n}_{\mu\nu}
B^{n}_{\alpha\beta}]\label{efectiveaction4d}
\end{eqnarray}
provided the internal components of the KR field $\chi^{n}(r)$ satisfy,
\begin{equation}
 -\partial^{2}_{r}\chi^{n}(r)=m_{n}^{2}\chi^{n}(r)h(r)\label{krdifeq1}
\end{equation}
along with  the orthonormality condition:
\begin{equation}
 \frac{1}{r_c}\int^{r_{max}}_{r_0}h^{3/4(r)}\chi^{m}(r)\chi^n(r)dr=\delta^{mn}\label{ortho1}
\end{equation}
Here $m_n$ represents the KK mass modes of KR field.\\
Let us now consider the massless mode of the KR field which is related to the
spacetime torsion. Eq.(\ref{krdifeq1}) yields the solutions
for the internal component $\chi^{0}(r)$ of KR massless mode as,
\begin{equation}
 \chi^{0}=\sqrt{2r_c}\frac{r_0}{R^{3/2}}\bigg(1-\frac{r_{0}^{2}}{r_{max}^{2}}\bigg)^{-1/2}\label{wavefunction1}
\end{equation}
Using eqs.(\ref{stringparameters1}) and eq.(\ref{warpfactor3}) in eq(\ref{wavefunction1}),
we obtain the  wavefunction for the massless mode of KR field in terms of the background fluxes on the IR 
brane located at $r_0$ as,
\begin{equation}
 \chi^{0}=\sqrt{2}~e^{-\pi K/g_s M}\bigg[\frac{r_{max}}{r_0}
\bigg(1+\frac{r_{max}}{r_0}\bigg)^{-1/2}\bigg]\label{wavefunction2}
\end{equation}
In \cite{ssgkr}, it was shown that in the background of the RS warped geometry, the zeroth mode of the KR field
(which is related to the spacetime torsion in 4-D spacetime) is heavily suppressed on the 
low energy brane.
This results into a heavy suppression of it's couplings to the 
brane fields (or SM fields) in comparison to the corresponding
couplings of the massless 4-D graviton.\\
 Here in string compactification, eq.(\ref{wavefunction2}) 
is again exhibiting an exponential suppression of the massless KR field
on the IR region and this suppression depends on the background flux parameters. \\
We now determine the coupling of the torsion field with 
the standard model (SM) spin-1/2 fermion fields, which are confined
to the low energy brane located at $r=r_0$ in the IR region. The interaction Lagrangian of the 
massless KR field and the SM fermions is,
\begin{equation}
\mathscr{L}_{\psi\overline{\psi}H^{0}}=\frac{1}{M^{3/2}}\overline{\psi}
  \left[ i\gamma^{\mu}\sigma^{\nu\lambda}H^{0}_{\mu\nu\lambda}(x^{\mu})\frac{\chi^{0}(r)}
{\sqrt{r_c}}\right] \psi\label{couplingzero1}
\end{equation}
Integrating over the extra dimensional part, the effective 4-D
Lagrangian becomes,
\begin{equation}
\mathscr{L}_{\psi\overline{\psi}H^{0}}=i\overline{\psi}\gamma^{\mu}\sigma^{\nu\lambda}
\bigg[\frac{e^{-4\pi K/3g_s M}}{M_{Pl}}\bigg(\frac{r_{max}}{r_0}\bigg)\bigg]H^{0}_{\mu\nu\lambda} \psi\label{couplingzero2}
\end{equation}
where we use the relation between the fundamental Planck scale $M$ and the 4-D Planck scale $M_{Pl}$
as,
\begin{equation}
M_{Pl}=\frac{M^{3/2}}{\sqrt{2R}}r_{max}\bigg(1-\frac{r_{0}^{2}}{r_{max}^{2}}\bigg)^{1/2}\label{mvsmpl}
\end{equation}
Eq.(\ref{mvsmpl}) is obtained by integrating the five dimensional 
gravitational action with respect to the internal component $r$
and comparing the derived effective 4-dimensional action with 
the standard 4-dimensional gravitational action.
We have also used eq.(\ref{warpfactor3}) in order to obtain the final 
expression ({\it {i.e}} eq.(\ref{couplingzero2})) 
for the coupling of the massless mode of the KR  field with the SM fermion fields on the IR brane.
It is now clear from eq.(\ref{couplingzero2}), that the coupling of the massless KR field
with the SM fermions on the IR brane is heavily suppressed by an exponential factor which depends on 
the background fluxes.\\
\subsection{\small Estimation of the background fluxes}\label{background fluxes}
Turning our attention to the experimental 
evaluation of four dimensional torsion components we see that in recent past 
the authors of \cite{cane,heckel,russel} have considered a possible 
violation of CPT and local Lorentz invariance due to the presence of
nonzero spacetime torsion in four dimensions. They have shown that torsion components
can be constrained to have an upper bound of the order of $10^{-31}$ GeV.
In this work, these experimental results \cite{cane,heckel,russel}
have been used to put bounds on various parameters 
of the model.\\
Equating the upper bound  $10^{-31}$ GeV on the 4-dimensional torsion 
component as evaluated in \cite{cane,heckel,russel} 
with the effective torsion component 
in the background of the KS geometry (see section\ref{KR field}), we obtain 
the lower bound on the ratio of the background fluxes and string coupling constant to be
$\frac{K}{g_S M}\sim 8.8$.\\
With this value for $\frac{K}{g_S M}$, eqn(\ref{warpfactor3}) yields a warp factor
$\sim 10^{-10}$. In order to address the gauge hierarchy problem 
in this background geometry, the value of $\frac{K}{g_S M}\sim 17.5$ such that
the warp factor $\sim 10^{-16}$. For this choice of the parameter $\frac{K}{g_S M}$, 
coupling becomes much more weaker which in turn reduces the possibility
of having signatures of spacetime torsion in future experiments even more.\\
 In summary this result depicts that in the background of the warped throat-like compactification 
the massless mode of the second rank antisymmetric KR field ans it's couplings are 
heavily suppressed on the IR region of the spacetime.
\section{Analysis of massless higher rank antisymmetric 
tensor fields in the Klebanov-Strassler geometry }\label{higher rank field}
Here in the background of the KS throat geometry, we investigate the behaviour of
massless higher rank antisymmetric tensor excitations of closed strings
on the low energy /IR brane.\\
We consider a rank-3 antisymmetric tensor field,
$X_{MNA}$, with the corresponding field strength tensor $Y_{MNAB}$.
The five dimensional action of $X_{MNA}$ can be written as,
\begin{equation}
 S=\int d^{5}x \sqrt{-G} Y_{MNAB}Y^{MNAB}\label{actionyads}
\end{equation}
G is the determinant of the 5 dimensional $AdS_5$ metric. 
In general for a rank $n$ antisymmetric gauge field $X_{a_{1}a_{2}....a_{n}}$,
one should be able to write down a rank-$(n+1)$ antisymmetric
field strength tensor as,
\begin{equation}
 Y_{a_{1}a_{2}....a_{n+1}}=\partial_{[a_{n+1}} X_{a_{1}a_{2}....a_{n}]}\label{yads1}
\end{equation}
Using the explicit form of the five dimensional metric (eq.(\ref{ksmetric})), the gauge 
fixing condition for the tensor field $X$ ({\it{i.e}} $X_{\mu\nu r}=0$) and 
the KK decomposition of the field X, 
we obtain the effective 4-dimensional action for the tensor field $X$,
\begin{eqnarray}
S_{x}=\int d^{4}x \sum_{n}\bigg[\eta^{\mu\lambda}\eta^{\nu\rho}\eta^{\alpha\gamma}\eta^{\beta\delta}Y_{\mu\nu\alpha\beta}^{n}
Y_{\lambda\rho\gamma\delta}^{n}\\\nonumber
+ 4 m_{n}^{2}\eta^{\mu\lambda}
\eta^{\nu\rho}\eta^{\alpha\delta} X_{\mu\nu\alpha}^{n}X_{\lambda\rho\gamma}^{n}\bigg]\label{action4yads}
\end{eqnarray}
where $m_{n}^{2}$ is defined through the relation,
\begin{equation}
-\partial_{r}^{2}\chi^{n}(r)=m_{n}^{2}h(r)\chi^{n}(r)\label{wavefunction3}
 \end{equation}
and $\chi^{n}$ satisfies the orthonormality condition,
\begin{equation}
\frac{1}{r_c}\int_{r_0}^{r_{max}} h^{5/4}(r)\chi^{m}(r)\chi^{n}(r)dr=\delta_{mn}\label{ocyads}
\end{equation}
Proceeding similarly as in case of the KR field in the preceding section, we obtain the wavefunction for the massless
mode of the tensor field $X$ on the IR brane located at $r=r_0$,
\begin{equation}
\chi^{0}= 2~e^{-5\pi K/3g_s M}\bigg(\frac{r_{max}^{2}}{r_{0}^{2}}\bigg)\bigg[1+\frac{r_{max}^{2}}{r_{0}^{2}}\bigg]^{-1/2}
\bigg(1+\frac{r_{max}}{r_0}\bigg)^{-1/2}\label{wavefunction4}
\end{equation}
We evaluate the interaction Lagrangian of the massless mode
 of the third rank antisymmetric tensor gauge fields
with the SM fermion fields on the low energy brane, located at $r=r_0$ as,
\begin{eqnarray}
\mathscr{L}_{\psi\overline{\psi}Y^{0}}&=&i\overline{\psi}\gamma^{\alpha}\Sigma^{\mu\nu\lambda}
~\sqrt{2}\bigg(\frac{e^{-2\pi K/g_s M}}{M_{Pl}}\bigg)\bigg(\frac{r_{max}^{2}}{r_{0}^{2}}\bigg)\\ \nonumber
&&\bigg[1+\frac{r_{max}^{2}}{r_{0}^{2}}\bigg]^{-1/2}
Y^{0}_{\mu\nu\alpha\beta}(x^{\mu})\psi
\label{couplingY1}
\end{eqnarray}
Thus, in the background of the KS throat geometry, the massless mode
of the higher rank tensor gauge field and it's
coupling with the SM fermions on the IR
brane is even more suppressed than the massless mode of the 
second rank antisymmetric KR field.\\
This analysis can easily be extended to even higher rank antisymmetric gauge fields 
which exhibits even higher suppression factor for all of them. 
\section{Conclusion}\label{conclusion}
String theory predicts the presence of massless modes of various
higher rank antisymmetric tensor fields. An inevitable question that arises is, 
why all these massless modes of these tensor fields are invisible in our Universe?
Here we try to address this question in the background of a warped throat-like KS 
string compactification. We consider various higher rank tensor fields 
in the background of the KS geometry and analyse the massless modes of
these tensor fields projected on our brane.\\
 Our result reveals that the massless modes of 
all these higher rank tensor gauge fields are heavily suppressed on the IR region of the
spacetime which can be depicted as the SM brane/our Universe.
 In case of the KR field which also can act as the spacetime torsion,
is exponentially suppressed on the SM brane, where the amount of suppression is parametrised 
by the background fluxes of the spacetime. We also show that the 
higher rank antisymmetric tensor gauge fields are even more suppressed than the
rank two antisymmetric KR field. In the context of the rank two antisymmetric
tensor field which is also interpreted as the spacetime torsion,
 we estimate the lower bound on the ratio of background fluxes
and string coupling constant by incorporating the 
experimental upper bound on the 4-dimensional torsion component.\\
Looking back to earlier works, where all the higher rank tensor gauge fields have been
analysed in the context of the warped RS scenario in five dimensions, it can be stated
in general that a warped nature of spacetime, in a string-inspired 
scenario can also explain the absence of any signatures of higher rank
antisymmetric tensor fields in our Universe for appropriate choices of background fluxes.

\end{document}